\pdfoutput=1
\documentclass{sigchi-ext}
\usepackage{hyperref}
\usepackage[T1]{fontenc}
\usepackage{textcomp}
\usepackage[scaled=.92]{helvet} 
\usepackage{graphicx} 
\usepackage{balance}  
\usepackage{booktabs} 
\usepackage{ccicons}  
\usepackage{ragged2e} 

\def\plaintitle{SIGCHI Extended Abstracts Sample File: Note Initial
  Caps} 
\def\emptyauthor{}
\def\plainkeywords{Human-Vehicle Interactions; Computer Vision; Ergonomics.}

\title{Building BROOK: A Multi-modal and Facial Video Database for Human-Vehicle Interaction Research}

\numberofauthors{6}
\author{%
  \alignauthor{%
    \textbf{Xiangjun Peng}\\
    \affaddr{University of Nottingham}  \\
    \affaddr{Ningbo, Zhejiang, China}\\
    \email{zy22056@nottingham.edu.cn} }\alignauthor{%
    \email{} } \vfil \alignauthor{%
    \textbf{Zhentao Huang}\\
    \affaddr{University of Nottingham}\\
    \affaddr{Ningbo, Zhejiang, China}\\
    \email{scyzh3@nottingham.edu.cn} }
    \alignauthor{%
    \email{} } 
    \vfil \alignauthor{%
    \textbf{Xu Sun}\\    
    \affaddr{University of Nottingham}\\
    \affaddr{Ningbo, Zhejiang, China}\\
    \email{xu.sun@nottingham.edu.cn} }
    \alignauthor{%
    \email{}}}

\definecolor{linkColor}{RGB}{6,125,233}
\hypersetup{%
  pdftitle={\plaintitle},
  pdfauthor={\emptyauthor},
  pdfkeywords={\plainkeywords},
  bookmarksnumbered,
  pdfstartview={FitH},
  colorlinks,
  citecolor=black,
  filecolor=black,
  linkcolor=black,
  urlcolor=linkColor,
  breaklinks=true,
}

\begin{document}

\CopyrightYear{2020}
\setcopyright{rightsretained}
\conferenceinfo{CHI'20,}{April  25--30, 2020, Honolulu, HI, USA}
\isbn{978-1-4503-6819-3/20/04}
\doi{https://doi.org/10.1145/3334480.XXXXXXX}
\copyrightinfo{\acmcopyright}

\maketitle

\RaggedRight{} 

\begin{abstract}
  With the growing popularity of Autonomous Vehicles, more opportunities have bloomed in the context of Human-Vehicle Interactions. However, the lack of comprehensive and concrete database support for such specific use case limits relevant studies in the whole design spaces.
  
  In this paper, we present our work-in-progress BROOK, a public multi-modal database with facial video records, which could be used to characterise drivers' affective states and driving styles. We first explain how we over-engineer such database in details, and what we have gained through a ten-month study. Then we showcase a Neural Network-based predictor, leveraging BROOK, which supports multi-modal prediction (including physiological data of heart rate and skin conductance and driving status data of speed) through facial videos. Finally we discuss related issues when building such a database and our future directions in the context of BROOK. 
  
  We believe BROOK is an essential building block for future Human-Vehicle Interaction Research. More details and updates about the project BROOK is online at \url{https://unnc-ucc.github.io/BROOK/}.
\end{abstract}


\printccsdesc

\section{Introduction}

  \marginpar{\textbf{Data Release} \\ The current version of BROOK would be released upon approached by research groups with ethical approval, please contact at scyzh3@nottingham.edu.cn.}
  
  \begin{marginfigure}[3pc]
\begin{minipage}{\marginparwidth}
    \centering
    \includegraphics[width=0.9\marginparwidth]{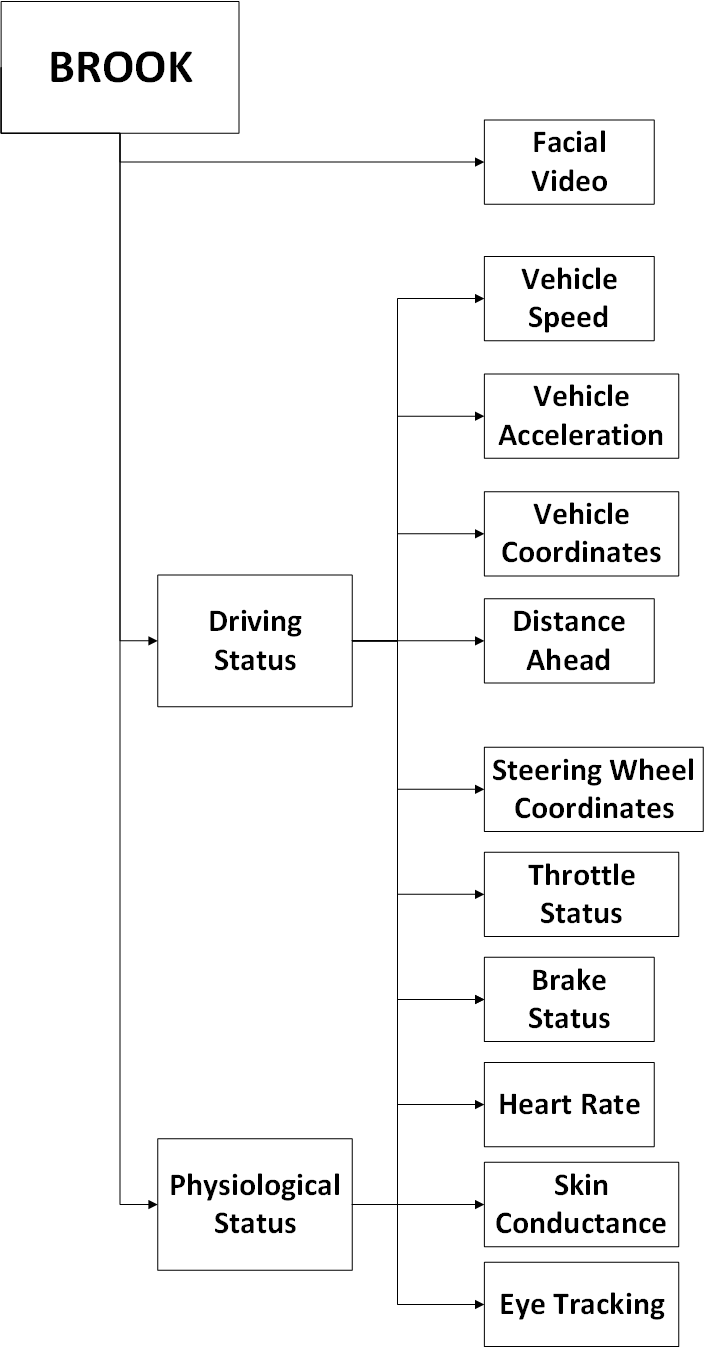}
    \caption{The Composition of the Current BROOK Database.}~\label{structure}
  \end{minipage}
\end{marginfigure}
  
  While autonomous vehicles are coming into our daily life and the work on Human-Vehicle Interaction (HVI) steadily progresses, \textbf{the question of missing specific data to support design the interaction between driver and the automated vehicle remains largely unsolved}. This would heavily limit future research prototypes and validations. Such limitation would even be amplified, when emerging Machine Learning techniques to improve HVI.
  
  To this end, BROOK\footnote{The name "BROOK" refers to multiple water flows, which stands for the key idea of our database} was established, as the first attempt, to address and tackle such issue. The goal of BROOK is to over-engineer a multi-modal and facial video database, which aims to support designs and research in HVI. The key idea, behind BROOK, is to \textbf{map as many kinds of data flows as possible in the same timeline}, when involved users are under the driving scenarios. Such guiding philosophy allows BROOK to support different research purposes, by selecting different types of data as the sources.
  
  In this paper, we report our past ten-month study as the first step of BROOK, which we collected real-world users' data from 34 drivers in 11 dimensions. Later, we developed 3 Neural Network-driven predictors for three specific categories, to showcase the power of BROOK database. Finally, we discuss several reflections from this study and identify further directions of this work.

  This paper made the following three major contributions:  \vspace{-1.5em}
  
  \begin{itemize}[leftmargin  = -1pt]
      \vspace{-0.2em}
      \item We present our work-in-progress BROOK, a multi-modal and facial video database, supported by customized driving scenarios. The current version of BROOK involves 34 drivers and their statistics (as structured in Figure \ref{structure}).
      \vspace{-0.2em}
      \item We showcase one potential use case of BROOK. Leveraging BROOK, we develop three Neural Network-driven predictors for heart rate, skin conductance and driving speed.
      \vspace{-2.0em}
      \item We quantitatively address the key limitation while developing BROOK, and discuss promising research directions of Human-Vehicle Interaction, with support from BROOK.
  \end{itemize}
 
\vspace{-1.5em}
  
\section{Background and Motivation}

\underline{\textbf{Human-Vehicle Interaction}} (HVI) is a growing sub-area of Human-Computer Interactions. Especially, with Autonomous Vehicles coming into daily life, the demand for intelligent and personalized HVI systems becomes intense. Since in-vehicles drivers start to become available from performing necessary driving behaviors. For example, classifications of Driving Styles, in long term, have started to be studied for intelligent and personalized HVI \cite{DS1,DS2,DS3,DS4,DS5}. 

\underline{\textbf{Data Support for HVI Research}} appears to be a missing part, especially when emerging techniques from other disciplines. As an essential part, domain-specific data support have bred a large number of advances in Data Science and Computer Vision, e.g. ImageNet \cite{ImageNet}, CIFAR \cite{CIFAR-10, CIFAR-online}, Kinetics \cite{Kinetics} and DEAP \cite{DEAP}. All those databases have supported the validations of different designs/prototypes and served as the source information for system developments.

\underline{\textbf{Our Vision}} is that, with the growing importance of Human-Vehicle Interaction research, a publicly available and over-engineered database is essential to support designs of interactive HVI. In response, we are building BROOK, a multi-modal and facial video database, which involves real-world drivers during customized driving scenarios. BROOK consists of both manual and autonomous driving behaviors, with 6 sensors and facial video records, from 34 drivers for around 20 minutes.

\newpage

\section{Overview}

\begin{margintable}[-2pc]
  \begin{minipage}{\marginparwidth}
    \centering
    \begin{tabular}{c c}
      {\small \textbf{Data Types}}
      & {\small \textbf{Size}} \\
      \toprule   
       Facial Video & 273 GB \\
      \midrule
       Heart Rates & 13.1 MB \\
      \midrule
       Skin Conductance & 48.3 MB \\
      \midrule
       Eye-tracking & 27.2 GB \\  
      \midrule
       Driving Status & 60.9 MB \\
      \bottomrule
    \end{tabular}
    \caption{A Summary of the Current BROOK Database.}~\label{BROOKOverview}
  \end{minipage}
\end{margintable}

\marginpar{\textbf{\underline{System Configuration}} \\ We host our driving simulator on a machine with two Intel Xeon E3-1225 processors and 32GB of RAM, with a NIVIDA RTX2070 GPU to ensure fluently visible scenarios. \vspace{10pt}}

\begin{margintable}[0pc]
  \begin{minipage}{\marginparwidth}
    \centering
    \begin{tabular}{c c c}
      {\small \textbf{}} & {\small \textbf{\# of Lanes}}
      & {\small \textbf{\# of Cars}} \\
      \toprule   
       1st & 4 & 32\\
      \midrule
       2nd & 4 & 60\\
      \midrule
       3rd & 8 & 60\\
      \midrule
       4th & 8 & 92\\
      \bottomrule
    \end{tabular}
    \caption{Configurations of Used Driving Scenarios and Scenes.}~\label{RoadPatterns}
  \end{minipage}
\end{margintable}

\begin{marginfigure}[0pc]
\begin{minipage}{\marginparwidth}
    \centering
    \includegraphics[width=0.9\marginparwidth]{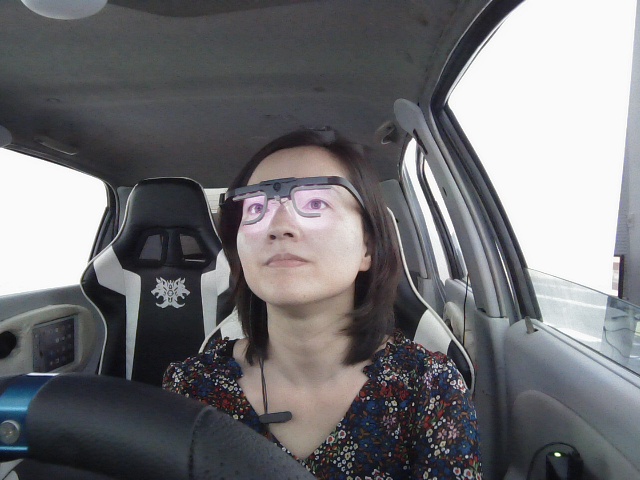}
    \caption{An Example of Data Collection Procedure.}~\label{exampleexperiment}
  \end{minipage}
\end{marginfigure}

In this section, we provide an overview of BROOK database. The current version of BROOK consists of 34 drivers, who has driven for around 20 minutes in automated and manual modes separately. BROOK have covered their facial videos, multi-modal data and driving status. Table \ref{BROOKOverview} provides key characteristics of the current BROOK version.  

Now BROOK consists of 11 dimensions of data, as shown in Figure 1, including: \textbf{Facial Video}, \textbf{Vehicle Speed}, \textbf{Vehicle Acceleration}, \textbf{Vehicle Coordinate}, \textbf{Distance of Vehicle Ahead}, \textbf{Steering Wheel Coordinates}, \textbf{Throttle Status}, \textbf{Brake Status}, \textbf{Heart Rate}, \textbf{Skin Conductance} and \textbf{Eye Tracking}. All data streams are consolidated with each other in the same time flow.

\section{Hardware and Software Support}

In this section, we provide details around hardware and software support, which has been utilized to build BROOK.

\textbf{\underline{Hardware Support.}} We deploy an ultra-clear motion tracking camera (for facial video), a heart rate sensor (for heart rates), a leather electrical sensor (for skin conductance) and a Tobii eye tracker (for Eye-tracking) for participants. 

Also, in order to collect detailed driving status information, we have integrated pressure sensors for Throttle and Brake, and we have ported Steering Wheel to obtain spatial changes from drivers during the whoe period.

\textbf{\underline{Software Support.}} We use OpenDS \cite{OpenDS,OpenDS-Official}, a open-source cognitive driving simulator, for driving scenario generations. Also, we utilise ErgoLAB \cite{ErgoLab} to coordinate sensor data output for storage. Finally, we build an in-framework tool to stream out speed and coordinates as driving status. 

\textbf{\underline{Region-based Customization.}} We customise our scenarios \cite{OpenDS-Scenarios} and scenes \cite{OpenDS-Scenes} based on the common patterns of roads in China, as shown in Table \ref{RoadPatterns}. This is because all participants during this study have held issued license from it, where they accumulated their driving experiences. 
\section{Procedure}
In this section, we illustrate the whole procedure (about 8 months) while collecting BROOK database. We first prepare necessary resources and tests to ensure the bulk of study run smoothly. Then we performed the bulk of the study to collect necessary data streams, in various different settings and modes separately.

\textbf{\underline{Preparations.}} Participants were first briefed about the study and were screened by a questionnaire to ensure that they were at low risk of motion sickness while experiencing the driving simulation. After the introduction, participants were given five minutes to sit in the simulator by themselves to adapt to the environment. The study only began when we ensured that participants understood the brief and were comfortable with the simulator.

\textbf{\underline{Data Collection.}} The experiment included three driving conditions, as explained in previous sections. Each participant began with the baseline manual mode, with drivers in control of the vehicle operations, following which they undertook the two autopilot experimental drives (standard and personalised versions\footnote{Determined by participants' behaviors in manual mode.}) which were completed in a counterbalanced order. During the study, driving data was logged. After each driving session, the participants were asked to fill out a questionnaire about the cognitive aspects of their experience, including an assessment of perceived trust, comfort and situational awareness. The study lasted approximately one hour for each participant.

\section{Case Study}

\begin{margintable}[0pc]
  \begin{minipage}{\marginparwidth}
    \centering
    \begin{tabular}{c c}
      {\small \textbf{Parameters}}
      & {\small \textbf{Value}} \\
      \toprule   
       Depth/Layers & 100 \\
      \midrule
       Growth Rate & 12 \\
      \midrule
       Dense Blocks & 4 \\
      \midrule
       Compression Factor & 0.5 \\
      \midrule
       Batch Size & 128 \\
      \midrule
       Initial Learning Rate & 0.1\\  
      \midrule
       Training Epochs & 50 \\
      \bottomrule
    \end{tabular}
    \caption{Pivotal Parameters about DenseNet for BROOK in details.}~\label{DenseNetDetails}
  \end{minipage}
\end{margintable}

\begin{marginfigure}[0pc]
\begin{minipage}{\marginparwidth}
    \centering
    \includegraphics[width=0.9\marginparwidth]{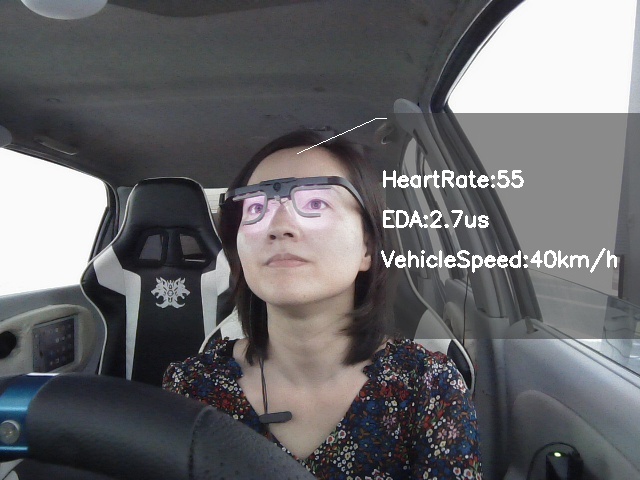}
    \caption{A Demonstrated Example of Our Prototype.}~\label{EstimatorExample}
  \end{minipage}
\end{marginfigure}

\begin{marginfigure}[0pc]
\begin{minipage}{\marginparwidth}
    \centering
    \includegraphics[width=0.9\marginparwidth]{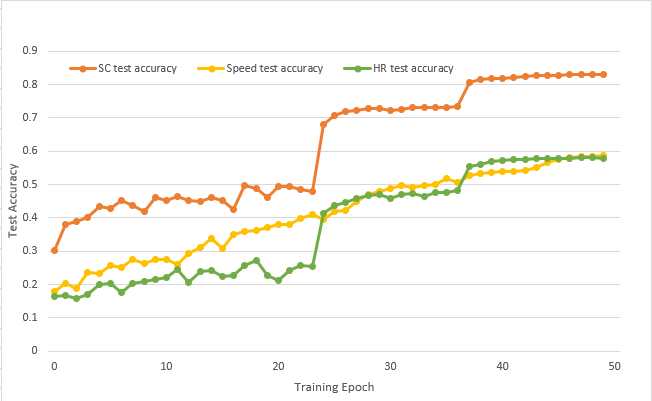}
    \caption{An Overview of Test Accuracy between skin conductance (Orange), heart rates (Green) and Vehicle Speed (Yellow).}~\label{Accuracy}
  \end{minipage}
\end{marginfigure}

In this section, we present an in-vehicle real-time design to estimate driver's multi-modal states (skin conductance and heart rates) and driving status (speed) through their facial expressions only. Powered by BROOK, we present three representatives of our trained models, which are specialised for these three kinds of data flows separately.

\textbf{\underline{Data Pre-processing.}} Before model training, we manually went through BROOK, recorded all "abnormal" \footnote{"abnormal" refers to those out-of-the-common-range data, which shall be within 60-100 (per minutes) for Heart Rates, 0.0-20.0 (uS) for Skin Conductance and 0-120 (km/h) for Vehicle Speed. And all of "abnormal" issues are confirmed through our video records.} data and exclude noisy data, which are caused by the unconscious operations during the setup and abort stages of our study. In total, the ready-to-train BROOK contains 320,000 frames/images, and we split the training, test and validation sets in a ratio of 8:1:1. 

\textbf{\underline{Methodology.}} To map data with frames, we simply label every frame of facial video streams with relevant data streams, which are Heart Rates (per minutes), Skin Conductance (uS) and Vehicle Speed (km/h) respectively. For each kind of data flows, we apply DenseNet as the model architecture,  since it saves many efforts of hyper-parameter adjustments through feature-map concatenation \cite{DenseNet}. 

In addition, there are two points to highlight. For those "abnormal" data, we re-assigned their labels with the lower-bound values, which are 60 (for Heart Rates), 20.0 (for Skin Conductance) and 120 (for Vehicle Speed). 

To support the whole pipeline, we utilise a face detector, extended from OpenCV \cite{OpenCV} to crop only the face from camera captures, as the input for our trained models.  

\textbf{\underline{Data Training.}} We trained BROOK on a NVIDIA GeForce GTX 1660 Ti GPU, and details about DenseNet architecture are provided in Table \ref{DenseNetDetails}. The majority of our settings are based on the recommended settings from DenseNet. Such design choice is because BROOK varies from all other commonly used data sets for existing neural network architectures, which are initially purposed for image classifications (e.g. CIFAR \cite{CIFAR-10,CIFAR-online}), and we hope to remain its effectiveness as much as possible. 

However, we did some lightweight adjustments of the training epoch, batch size and some other hyper-parameters which would improve the accuracy. For our case, the initial learning rate is set to 0.1, and is divided by 10 at 50\% and 75\% of the total number of training epochs.

\textbf{\underline{Showcase.}}
As Figure \ref{EstimatorExample} shown, our estimation prototype reflects all estimated results directly. The backbone of this prototype is three different models, which are for three different kinds of data flows.

\textbf{\underline{Accuracy.}}
We also evaluated our prototype through validation set, in terms of accuracy. The single-crop validation accuracy is reported in Figure \ref{Accuracy} as a function of training epochs. 
The validation accuracy of skin conductance, heart rates and vehicle speed are 82.96\%, 57.90\% and 58.75\% respectively. Our estimation on Skin Conductance is better than other data significantly, it might be caused by the smaller timeslot among data streams, which substantially provides more valuable information during training. 

One thing to be highlighted, is that initial hyper-parameter settings are mainly optimized for image classifications. We are confident that more studies through extensive hyper-parameter adjustment could improve the accuracy of our prototypes in the end.

\begin{figure*}
  \centering
  \includegraphics[width=1.3\columnwidth]{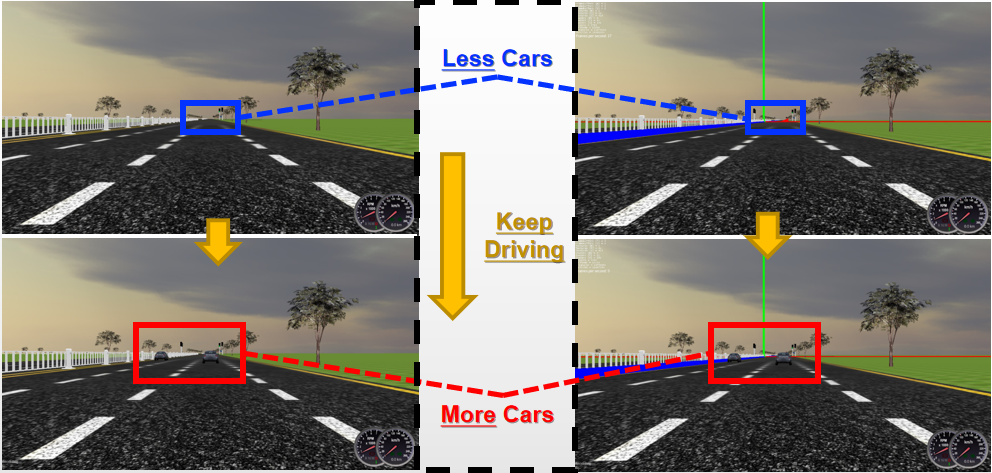}
  \caption{An Illustration of Our Case Study to Address the Frequency Flip Issue. The right series are equipped with the detailed monitor to record the frequency.}~\label{DrivingDensity}
\end{figure*}

\begin{marginfigure}[-20pc]
\begin{minipage}{\marginparwidth}
    \centering
    \includegraphics[width=0.9\marginparwidth]{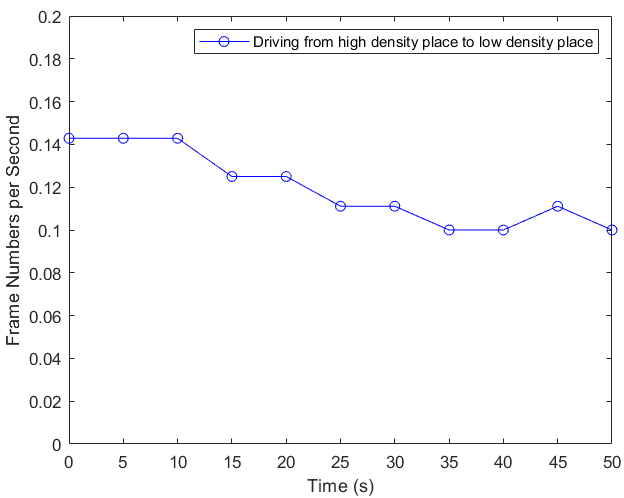}
    \caption{Visual Trend of Frequency Changes through the whole procedure, as shown in Figure 5.}~\label{fig:marginfig}
  \end{minipage}
\end{marginfigure}

\begin{margintable}[0pc]
  \begin{minipage}{\marginparwidth}
    \centering
    \begin{tabular}{c c}
      {\small \textbf{Timespots}}
      & {\small \textbf{Frequency}} \\
      \toprule   
       5 & 0.142 \\
       15 & 0.124 \\
      \midrule
       15 & 0.124 \\
       25 & 0.113 \\
      \midrule
       25 & 0.113 \\
       35 & \textbf{\underline{0.096}} \\
      \midrule
       35 & \textbf{\underline{0.094}} \\
       45 & 0.119\\  
      \bottomrule
    \end{tabular}
    \caption{Detailed Breakdown of Frequency Changes at Different Time Spots, as the procedure shown in Figure 5.}~\label{tab:table2}
  \end{minipage}
\end{margintable}

\vspace{-3.5em}
\section{Key Limitation}

In this section, we address the key limitation, based on our ten-month preliminary tryouts to build and utilize BROOK. We performed a case study to illustrate our concerns in details, which are presented as follow. 

\textbf{\underline{Motivation.}} During our studies among different Traffic Scenarios, we found that the frequency of frames, in our driving simulator, might drift while the density of the scenarios changes. Therefore, to quantitatively address this issue, we performed a case study to observe how severe this issue is and whether it would endanger the performance of our Vehicle Speed predictor further.

\textbf{\underline{Scenario Designs.}} We implemented a straight road, with several vehicles coming near. Also, we set our vehicle (in the first-person perspective) at a constant speed to go forward. Figure 5 illustrates the whole procedure in details, and it automatically ends when two vehicle sets have passed one another.

\textbf{\underline{Timespot Breakdown.}} We first visualize the relatively low-precision frequency during the whole process. As shown in Figure \ref{fig:marginfig}, the drift has been confirmed, and the frequency would become lower while there are more and more vehicles within the visual field.

To further quantify such issue, we deploy a high-precision frequency monitor for the overall procedure. As shown in Table 4, detailed drifts are quantified with different timespots. As highlighted, we draw the key observation that, when vehicles interleave within the visual range, the frequency would flip into another unit level.

\textbf{\underline{Why Important?}} Such flip could cause significant penalty for driving-status predictor, since all data are streamed out from the simulator and labelled frame-by-frame. In our case, we might have mislabelled much Vehicle Speed data in the first place, which caused accuracy degradation, since we added lots of vehicles for more realistic scenarios.

\section{Discussions}

We believe BROOK would be a building block for many future research, which aims to stretch the boundaries of theory and practice, in the context of Human-Vehicle Interaction (HVI). Hereby this moment, we summarize three aspects of potential research agenda, which are imposed with the support of BROOK.

\textbf{\underline{Accurate In-vehicle Face-to-Multimodal Predictors.}}\\
\vspace{-0.005em}

Tailed to the need for personalized HVI systems, multi-modal data support could bring a lot of benefits to assist decision-making procedures for different drivers. However, adapting multiple sensors for drivers/passengers is not user-friendly and hard to be integrated. We believe BROOK provides a unique series of opportunities to develop and validate more accurate Face-to-Multimodal predictors (e.g. more detailed facial expression \cite{FE1,FE2}) inside the vehicles, which has a great potential to be the data source for future HVI systems in practice.

\textbf{\underline{Long Short-Term Characterization of Drivers.}}\\
\vspace{-0.005em}

Current studies have paid a lot of attention on the characterizations and classification of driving styles, but few attempts to adapt the instant reflections from drivers/passengers. We believe BROOK provides offers a great many chances for taking in-depth analysis and modeling of drivers/passengers in both Long and Short Term (e.g. Long Short-Term Memory (LSTM) \cite{LSTM}). Despite driving styles from long-term inference, future HVI systems could be capable of providing real-time reflections on drivers'/passengers' affective states, to further improve its quality of service. 

\textbf{\underline{Novel Adaptions of Neural Network Models into HVI.}}\\
\vspace{-0.005em}

One of the key constraints, to adapt complicated decision-making procedures into Autonomous Vehicles, is the limitation of computationally power. Even for our demonstration, we faced tremendous performance challenges while training video-based data set and obtain preliminary results. Compression (e.g. \cite{Compression/ICLR16, Compression/ECCV18}), context-aware combination (e.g. \cite{CA1,CA2,CA3}) and reconfiguration (e.g. \cite{Configuration}) would all be useful in the context of HVI. It's clear that, even outside HVI community, BROOK is still quite valuable to various other communities, since the usage and training of video data are still in debts (e.g. \cite{DebtsVideo1,DebtsVideo2}) and need more sources to validate novel proposals. 

\section{Conclusion and Future Work}

We presented BROOK, a multi-modal and facial video database for Human-Vehicle Interaction Research. A ten-month study has merged 34 drivers into the current version of BROOK database, with manual and auto-pilot modes separately. Based on BROOK, we showcased a novel usage of Face-to-Multimodal predictor and evaluate it in details. Also, we have addressed the key limitation of BROOK project, and we briefly discuss the future research agenda, enlighten by BROOK.

We hope BROOK could stimulate more intellectual contributions to the growing Human-Vehicle Interaction field. Since BROOK is a long-term project, our future work would consist of multiple aspects for expansion of it, including collecting valuable data through more realistic simulation support \cite{Vid2vid}, improving simulator-related productivity \cite{Onerios, Ommniverse} and providing more variants of BROOK as needed. 

More details and updates about the project BROOK is online at \url{https://unnc-idl-ucc.github.io/BROOK/}.


\section{Acknowledgements}

We thanked for all members of User-Centric Computing Group and CHI'20 reviewers for their valuable and insightful feedback. This research is generously supported by LandRover and National Youth Fund No. 123456.

\bibliographystyle{SIGCHI-Reference-Format}
\bibliography{reference}

\end{document}